\documentclass[twocolumn, showpacs,superscriptaddress,nofootinbib,floatfix, amsfonts, aps]{revtex4}

\usepackage{amsmath}
\usepackage{bm}
\usepackage{ulem}
\usepackage{graphicx}
\usepackage{mathrsfs}
\usepackage{footmisc}
\usepackage{multirow}
\usepackage{graphicx}
\usepackage{booktabs, ctable}
\usepackage{xcolor, color, framed}

\graphicspath{{figs/}}
\begin{document}

\title{
Probe the tilted Quark-Gluon Plasma with charmonium directed flow}
\author{Baoyi Chen} 
\email{baoyi.chen@tju.edu.cn}
\affiliation{Department of Physics, Tianjin University, Tianjin 300350, China} 
\author{Maoxin Hu} 
\affiliation{Department of Physics, Tianjin University, Tianjin 300350, China} 
\author{Huanyu Zhang}  
\affiliation{Department of Physics, Tianjin University, Tianjin 300350, China} 
\author{Jiaxing Zhao}
\email{zhao-jx15@mails.tsinghua.edu.cn}
\affiliation{Department of Physics, Tsinghua University, Beijing 100084, China} 

\date{\today}

\begin{abstract}
Charmonium directed flows are studied based on transport model 
coupled with the realistic three dimensional 
expansions of the bulk medium. The non-central symmetric nucleus-nucleus collisions can 
generate the rotating quark-gluon plasma (QGP) 
with symmetry-breaking longitudinal distributions. In 
$\sqrt{s_{NN}}=200$ GeV Au+Au semi-central collisions, 
charmonium are primordially produced in the initial hard process, they are mainly 
dissociated by the initial tilted source with high temperatures and then  
move out of the bulk medium to keep the early 
information of the medium. The momentum distribution of primordially 
produced charmonia is less affected 
by the hydrodynamic expansions of QGP where its tilted shape 
is being diluted. This biased dissociation can generate directed flows of 
$J/\psi$ and $\psi(2S)$ which are much larger than the values of light charged 
hadrons and open heavy flavor. Charmonium directed flows can help to quantify 
the rapidity-odd distributions of QGP initial energy densities in nucleus-nucleus collisions. 
\end{abstract}


\maketitle

It's well known that a strong electromagnetic field and 
a strong vorticity field can be generated in the non-central relativistic 
heavy ion collisions. Due to the coupling between the parton spin 
and vorticity field, some interesting effects about the polarizations of 
quarks and hadrons have been studied~\cite{Liang:2004ph,Pang:2016igs, Becattini:2015ska}. 
Theoretical and experimental studies 
show that the light charged hadrons~\cite{Bozek:2010bi, Adamczyk:2014ipa, Nara:2016phs} 
and open heavy flavor such as D-mesons~\cite{Adam:2019wnk, Chatterjee:2017ahy} carry 
non-zero directed flows from the anisotropic expansions of the bulk medium. 
The magnitudes of these effects induced by the vorticity field depend sensitively on the 
rotations of initial QGP in semi-central nucleus-nucleus (AA) collisions.  
The evolutions of rotating QGP can be simulated by the hydrodynamic model. 
The initial energy density of QGP produced in the relativistic heavy ion collisions 
can be extracted with other models and constrained by the final distributions of charged 
particles.  Directed flows of charged particles and D mesons in experiments 
indicate 
the rapidity-odd energy density of QGP in the longitudinal direction. 
This tilted magnitude 
of the energy density will be diluted by the hydrodynamic expansions. 
Different from light partons, 
heavy quarks are produced in parton hard scatterings, and symmetrically distributed in 
forward-backward rapidities at the moment of nuclear collisions. 
In forward and backward rapidities of the tilted medium, 
heavy quarks and quarkonium undergo different number 
of elastic(inelastic) collisions with the bulk medium when moving in the different 
directions in the transverse plane. 
This results in the final anisotropic momentum distribution of heavy flavors 
in the transverse plane, such as directed flow $v_1$, 
elliptic flow $v_2$, etc. 
For open heavy flavors, 
their dragged movement by the QGP expansion 
also contributes to the anisotropic flows 
of charm quarks~\cite{Adamczyk:2014ipa, Adam:2019wnk}. The large uncertainty of 
heavy quark diffusion coefficient prevents further solid conclusions about D-meson 
directed flow~\cite{Dong:2019unq}.   

Charmonium as a bound state is first proposed to be a probe of the deconfined phase 
in the early stage of relativistic heavy ion collisions~\cite{Matsui:1986dk}. 
They are mostly dissociated in the initial stage where 
medium temperature is high. In the later 
stage of QGP hydrodynamic expansions, charmonium momentum distribution 
is not effectively changed by the elastic scatterings with light partons 
because of charmonium large mass and zero color charge. This makes 
charmonium a relatively 
clean probe for the initial anisotropic deposition of the bulk medium. 
Primordial charmonia are produced symmetrically in forward-backward rapidity with 
isotropic transverse momentum distribution in Au+Au collisions. After experiencing 
biased dissociations in tilted source, the rapidity-odd distributions and 
transverse elliptic distributions of the initial QGP result in the directed and elliptic 
flows of charmonia respectively.
These observables help to constrain the magnitude of the tilt in the  
longitudinal deposition of initial energy density in Au+Au collisions.

Charmonium evolution in the hot medium has been widely 
studied with rate equation~\cite{ Grandchamp:2003uw,Zhao:2007hh} and transport 
model~\cite{Zhu:2004nw, Yan:2006ve, Yao:2018nmy}. From the lattice QCD calculations, 
heavy quarkonium potential is partially screened 
by the deconfined medium. Charmonium bound states can also be dissociated through the 
inelastic scatterings with partons. Both of these hot medium effects suppress the 
charmonium final yields, and can be included in the transport equation, 
\begin{align}
\partial_t f_\psi({\bf x},{\bf p}, t) + {\bf v}\cdot {\bf{\bigtriangledown}} 
f_\psi({\bf x}, {\bf p}, t) 
= &- \alpha_\psi({\bf p}, T) f_\psi({\bf x}, {\bf p}, t)  \nonumber \\
& + \beta_\psi({\bf p}, T)
\label{trans-1}
\end{align} 
where ${\bf v}={\bf p}/E$ is the charmonium velocity. $T$ is the QGP local temperature 
depending on coordinates and time. Charmonium density in phase 
space $f_\psi({\bf x}, {\bf p}, t)$ changes with the time and coordinate due to their motions. 
$f_\psi$ can also be affected by hot medium effects 
labelled as $\alpha_\psi$ and $\beta_\psi$ terms 
on the right hand side of Eq.(\ref{trans-1}).  
$\alpha_\psi$ is charmonium dissociation rate in the QGP, 
the formula can be written as~\cite{Zhu:2004nw}, 
\begin{align}  
\alpha_\psi({\bf p},T) = {1\over E_T}\int {d^3{\bf k}\over (2\pi)^3 E_g}
\sigma_{g\psi}({\bf p}, {\bf k}, T) 
 F_{g\psi}({\bf p}, {\bf k}) f_g({\bf k},T)
\end{align}
where $E_T=\sqrt{m_\psi^2+p_T^2}$ is the transverse energy, $\bf k$ and $E_g$ is the gluon 
momentum and energy. $F_{g\psi}$ is the flux factor. 
The dissociation rate is proportional to gluon density $f_g$ and 
inelastic cross section $\sigma_{g\psi}$ between 
gluon and charmonium. The form of 
inelastic cross section for charmonium ground state $J/\psi$ in vacuum 
is calculated with the method of operator production expansion~\cite{Bhanot:1979vb}. 
Its value at finite temperature 
is obtained by reducing its binding energy in the formula to consider the color 
screening effect~\cite{Satz:2005hx}. The 
inelastic cross section 
for excited state ($\chi_c$, $\psi(2S)$) is obtained by the geometry scale with 
the ground state $J/\psi$, please see 
the details in Ref.\cite{Chen:2018kfo}.   
Meanwhile, within the deconfined phase, uncorrelated 
charm pairs produced in the hard process of initial hadronic collisions evolve randomly. 
They may recombine into a new bound state~\cite{Andronic:2006ky, Yao:2018sgn}, 
labelled with 
$\beta_\psi$ in Eq.(\ref{trans-1}). 
The regeneration term depends on the densities of charm and anti-charm quarks and their 
combination probability into a charmonium state~\cite{Shi:2017qep}. 
The regeneration contribution is suppressed in the 
non-central AA collisions at the RHIC energy due to the small density 
of charm pairs in QGP in these collisions~\cite{Liu:2009nb,Du:2015wha}. 
Therefore, we neglect the regeneration contribution in this work.

Charmonium primordial distribution in AA collisions can be 
obtained from the distribution in
pp collisions with the modification of cold nuclear matter 
effect~\cite{Lansberg:2016deg,Kusina:2017gkz,Emelyanov:1999pkc,Emelyanov:1998phs,McGlinchey:2012bp}. 
The primordial distribution in the transverse plane in AA collisions 
is written as~\cite{Chen:2016dke},  
\begin{align}
f_{t=0}({\bf x}_T, {\bf p}_T, y_p|{\bf b}) = &T_A({\bf x}_T^A) 
T_B({\bf x}_T^B) 
 {d^2 \sigma_{J/\psi}^{pp}\over dy_p 2\pi p_T dp_T}\times \mathcal{R}_{AB}
\end{align}
where ${\bf x}_T^A = {\bf x}_T+{{\bf b}\over 2}$ and 
${\bf x}_T^B = {\bf x}_T-{{\bf b}\over 2}$. 
The centers of two nuclei are located at ${\bf x}_T=(\pm b/2,0)$. 
$T_{A}({\bf x}_T)=\int_{-\infty}^{+\infty} dz_{A} \rho_{A}({\bf x}, z_{A})$ 
is the thickness function of nucleus A(B). Nuclear density $\rho_A$ is taken as Woods-Saxon 
distribution. 
$\mathcal{R}_{AB}$ is the 
modification factor from cold nuclear matter effect. We employ the 
EPS09 NLO model to parameterize the homogeneous shadowing factor for 
charmonium in the entire regions of $p_T$ and rapidity~\cite{Eskola:2009uj}, 
and take the 
 value $\mathcal{R}_{AB}=0.85$ in $\sqrt{s_{NN}}=200$ GeV 
Au+Au collisions. The shadowing effect is relatively weak at the RHIC colliding energy 
compared with the LHC colliding energies.  
It shows clear rapidity dependence in d+Au 
collisions, but tends to be flat with rapidity 
in symmetric AA collisions~\cite{Chen:2015iga}. 
As we employ the homogeneous factor for the shadowing effect, 
$\mathcal{R}_{AB}$ suppress the charmonium nuclear modification factor and does not 
affect the final momentum anisotropies according to their definitions. The detailed 
$p_T$-dependence in $\mathcal{R}_{AB}$ 
does not significantly change the $p_T$-integrated $v_1$ and $v_2$. 
Charmonium anisotropic flows are 
mainly generated by the interactions with the hot medium in AA collisions. 
The differential cross section in 
pp collisions 
as a function of transverse momentum $p_T$ and rapidity $y_p$ is parametrized 
from the experimental data with the formula, 
\begin{align}
{d^2\sigma^{pp}_{J/\psi}\over dy_p 2\pi p_T dp_T} &= {(n-1)\over \pi (n-2)\langle p_T^2\rangle }
[1+{p_T^2\over (n-2)\langle p_T^2\rangle}]^{-n} {d\sigma_{J/\psi}^{pp}\over dy_p} \\
{d\sigma_{J/\psi}^{pp}\over dy_p} &= Ae^{-By_p^2}{\rm{cosh}}(Cy_p)
\end{align}
with $n=6.0$, mean squared transverse momentum in the central rapidity is taken from the 
experimental data
$\langle p_T^2\rangle|_{y_p=0}=4.14$ $\rm{(GeV/c)^2}$~\cite{Adare:2006kf}, 
its rapidity dependence is fitted with 
$\langle p_T^2\rangle=\langle p_T^2\rangle|_{y_p=0}\times [1-(y_p/y_{pmax})^2]$ 
with the charmonium maximum rapidity defined as $y_{pmax}= {\rm{cosh}}^{-1}(\sqrt{s_{NN}}/(2E_T))$ 
in pp collisions. Parameters in the rapidity dependence of charmonium 
cross section is fitted to be $A=0.75$ $\rm{\mu b}$, $B=0.61$, $C=1.2$ with the 
data in forward rapidities~\cite{Adare:2006kf,Chen:2015iga}. 
Initial distributions of charmonium excited states are 
scaled from $J/\psi$ distribution with a factor measured in pp collisions. 
As nucleus is accelerated to the speed of $v_N\sim 0.9999c$, 
nuclear shape is strongly 
Lorentz contracted in the longitudinal direction, charmonium primordial 
longitudinal distribution is simplified as  
$f_{t=0}({\bf x}_T, z, {\bf p}_T, y_p|{\bf b}) =f_{t=0}({\bf x}_T,{\bf p}_T, y_p|{\bf b})
\delta(z)$. After nuclear collisions, 
QGP needs a period of time $\tau_0\sim 0.6$ fm/c 
to reach local equilibrium and to start transverse expansion. Both hydrodynamic models and 
transport equation 
Eq.(\ref{trans-1}) start from $\tau_0$. 
Before this time scale, charmonium is free streaming 
in the QGP pre-equilibrium stage. 
With primordial 
charmonia produced at the very beginning of  
nuclear collisions, their dissociations 
mainly happen in the early stage of QGP which makes them sensitive to the initial 
anisotropic distribution of QGP energy densities.

\begin{figure}[hbt]
\centering
\includegraphics[width=0.40\textwidth]{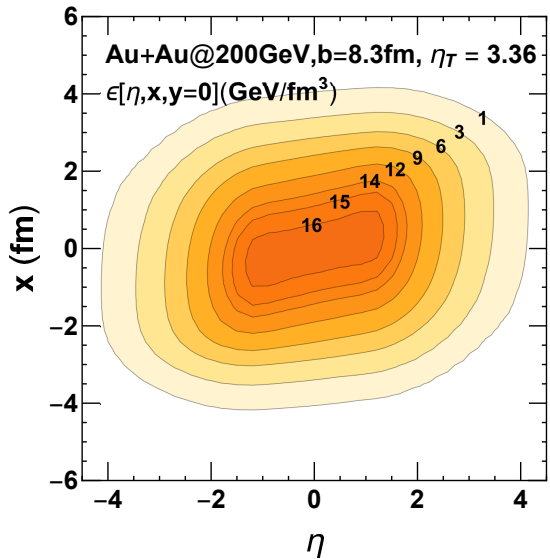}
\caption{ (Color online) 
Spatial distributions of QGP initial energy density in the $x-\eta$ plane in 
$\sqrt{s_{NN}}=200$ GeV Au+Au collisions with the impact parameter $b=8.3$ fm.  
The transverse coordinate is chosen as ${\bf x}_T = (x,0)$. The parameter 
for the QGP tilted magnitude is taken as $\eta_T=3.36$. Different colors in the 
figure represent different values ($1\sim 16$ $\rm{GeV/fm^3}$) 
of the QGP energy density. 
}
\label{lab-fig1}
\end{figure}

In order to calculate the charmonium directed and elliptic flows, 
one needs to generate realistic QGP evolutions at $\sqrt{s_{NN}}=200$ GeV Au+Au collisions. 
When two nuclei collide with each other, 
the nucleus moving with the positive (negative) 
rapidity tends to tilt the produced bulk medium to positive (negative) 
x with respect to the beam axis for positive (negative) $z$.  
To explain the rapidity spectra of charged particles in Au+Au collisions, 
the initial tilted entropy density $s(\tau_0, {\bf x}_T, \eta_\parallel)$ is 
parametrized with~\cite{Bozek:2010bi,Chatterjee:2017ahy}, 
\begin{align}
s(\tau_0, {\bf x}_T, \eta_\parallel) =& s_0
\times
\exp[-\theta(|\eta_\parallel|-\eta_\parallel^0) 
{(|\eta_\parallel|-\eta_\parallel^0)^2 \over 2\sigma^2}] \nonumber \\
&\times [ c_{hard} N_{coll} + (1-c_{hard}) 
(N_{part}^{+} \zeta_{+}(\eta_\parallel)  \nonumber \\
&+ N_{part}^{-} \zeta_{-}(\eta_\parallel))] 
\end{align} 
where $s_0$ and the exponential term represent initial entropy density and the 
rapidity distribution 
to reproduce the rapidity spectra of charged particles in AA collisions. 
$N_{part}^{+}$ and $N_{part}^{-}$ is the number of participants in the forward 
and backward rapidities respectively. $N_{coll}$ is the number of binary collisions. 
The initial entropy density comes from both soft ($N_{part}$ terms) 
and hard ($N_{coll}$ terms) collision process with a fraction $1-c_{hard}$ and 
$c_{hard}$ respectively. The parameters are fixed to 
be $c_{hard}=0.05$, $\eta_\parallel^0=1.0$, $\sigma=1.3$.  
The specific rapidity-odd distribution in entropy is introduced by 
the function $\zeta_{\pm}(\eta_\parallel)$. 
It's taken as $\zeta_{\pm}(\eta_\parallel) = {\eta_T \pm\eta_\parallel\over 2\eta_T} $ at 
$|\eta_\parallel|\le\eta_T$. 
For $|\eta_\parallel|>\eta_T$, they are fixed to their values at $\pm \eta_T$. 
$\eta_T$ controls the magnitude of the 
tilt of the initial QGP and is taken as $\eta_T=3.36$~\cite{Chatterjee:2017ahy}. 
Different values of parameter $\eta_T$ will be taken, 
to provide different tilted backgrounds for charmonium evolutions. 
QGP initial energy distribution is plotted in Fig.\ref{lab-fig1}. One can clearly see 
the tilt of bulk medium distribution in forward and backward rapidities, 
which will result in biased dissociations and directed flows for charmonia. 
QGP expansions are simulated with (3+1)D hydrodynamic model 
(MUSIC) 
followed by the crossover phase transition to the hadronic 
phase~\cite{Schenke:2010nt, Schenke:2010rr}. 
Its equation of state is 
obtained with the interpolation between lattice data and a
hadron resonance gas~\cite{Huovinen:2009yb}. In this work, 
hadron phase contribution on charmonium dissociation is neglected 
for simplicity.

Different from soft collisions, 
binary collision profiles are symmetric in forward and backward rapidities.  
Charmonium initial production from hard collisions is not tilted. 
Therefore, its distribution is shifted relative to the tilted bulk medium. As the tilted 
shape of the 
medium is more obvious in the early stage of QGP evolutions where 
charmonium dissociation effect is the strongest, charmonium bound states are 
sensitive to the early information of QGP compared with the observables such as 
light hadrons or open heavy flavors. The nuclear modification factors $R_{AA}$ of $J/\psi$ and 
$\psi(2S)$ based on this framework have been extensively studied and 
compared with the experimental data 
in Ref.\cite{Chen:2018kfo, Zhou:2014kka}.


\begin{figure}[ht]
\centering
\includegraphics[width=0.35\textwidth]{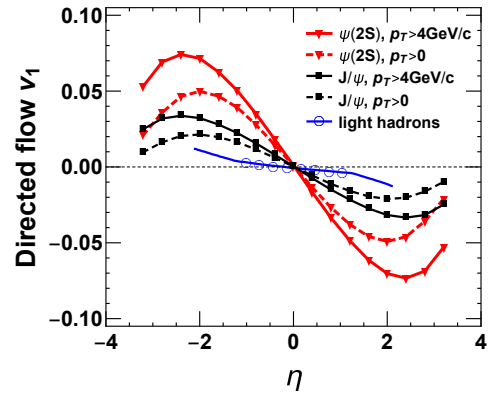}
\caption{(Color online) 
$J/\psi$ and $\psi(2S)$ directed flows 
as a function of $\eta$ in $\sqrt{s_{NN}}=200$ GeV Au+Au collisions. Charmonium 
$p_T$ range is integrated in $p_T>0$ an $p_T>4$ GeV/c respectively with dashed and 
solid lines. Impact parameter is taken as $b=8.3$ fm for the centrality 0-80\%.  
Charged hadron directed flow is also included for comparison: 
theoretical calculation with hydrodynamic model is cited from Ref.\cite{Chatterjee:2017ahy}, 
experimental data is from STAR measurements~\cite{Abelev:2008jga}. 
The parameter 
for the QGP tilted magnitude is taken as $\eta_T=3.36$. 
}
\label{labfig-v1}
\end{figure}

In Fig.\ref{labfig-v1}, charmonium directed flows integrated in different $p_T$ regions are 
compared with the charged particles. The impact parameter is set with $b=8.3$ fm for the 
centrality 0-80\%. 
In the figure, both $J/\psi$ (thin black line) and $\psi(2S)$ (thick red line)   
directed flows are several times larger than the value of charged particles measured by STAR, 
because charmonia are 
directly dissociated by the asymmetric source in the early stage where the tilt 
is the largest. Excited state $\psi(2S)$ with smaller binding energy 
is easier to be dissociated in QGP due to the larger $\sigma_{g\psi(2S)}$, 
compared with the ground state $J/\psi$.  
$\psi(2S)$ directed flow is larger than the $v_1$ of $J/\psi$ and charged particles. 
Note that even though 
the directed flows of both charmonia and charged hadrons are due to the tilted source 
in semi-central AA collisions, their processes are different. For charged particles and 
D mesons, their directed flows are mainly 
from the drag of tilted bulk medium with hydrodynamic 
expansions~\cite{Chatterjee:2017ahy}. Charmonium directed 
flows are dominated by the biased dissociations due to the difference of 
their path lengths in the QGP initial stage. 
In the later stage of QGP expansions, 
charmonium momentum distribution is less affected by the elastic 
collisions with QGP. For the lines with $p_T>0$ in Fig.\ref{labfig-v1}, charmonium 
directed flows are smaller. As charmonium 
with smaller transverse velocity will stay longer inside the QGP. 
Hydrodynamic expansions will reduce the tilt of 
QGP, which makes charmonium biased 
dissociations not so obvious. 
For those charmonium with large momentum, they 
experience the initial tilted QGP and move out of the bulk medium quickly to keep the 
early information of the tilted source. Charmonium $v_1$ can be 
enhanced by around 
50\% when $p_T$ range is shifted from $p_T>0$ to $p_T>4$ GeV/c.

\begin{figure}[ht]
\centering
\includegraphics[width=0.32\textwidth]{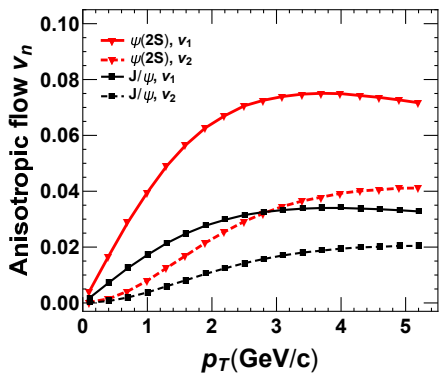}
\caption{ 
(Color online)  
Directed and elliptic flows of $J/\psi$ and $\psi(2S)$ as a function of transverse momentum 
$p_T$ with the fixed rapidity $\eta$=-2.4, 
in $\sqrt{s_{NN}}=200$ GeV Au+Au collisions. Other parameters are the same with 
Fig.\ref{labfig-v1}. 
}
\label{labfig-v2v1-pt}
\end{figure}

To show that charmonia with larger $p_T$ can keep more early information of the tilted source, 
we plot charmonium $v_1(p_T)$ in Fig.\ref{labfig-v2v1-pt}. The magnitude 
of charmonium $v_1$ 
increases with $p_T$. This is consistent with Fig.\ref{labfig-v1}.
As charmonia suffers 
different dissociations in positive- and negative- 
x-direction from the initial QGP due to the path-length-difference, 
and move out of the medium quickly, 
this makes charmonium $|v_1|$ and $v_2$ increase with $p_T$. This effect saturates at 
high $p_T$ and does not drop to zero, please see Fig.\ref{labfig-v2v1-pt}. 
The momentum anisotropy induced by this path-length-difference mechanism 
is different from the situations of light hadrons or D mesons which 
momentum anisotropies come from the 
QGP anisotropic expansion and usually satisfy the mass ordering: particles with 
smaller mass are easier to reach kinetic equilibrium with QGP expansion and carry 
larger collective flows.

The external fields such as electromagnetic fields can also contribute to the 
directed and elliptic flows of open charm quarks. However, $J/\psi$ and $\psi(2S)$ 
as a bound state of $c\bar c$ dipole, carry zero net electric and color charges.   
Its evolutions is less affected by the external fields, act 
as a relatively clean probe for the initial asymmetric distributions of the bulk medium 
in longitudinal direction and the 
transverse plane. 

\begin{figure}[!t]
\centering
\includegraphics[width=0.35\textwidth]{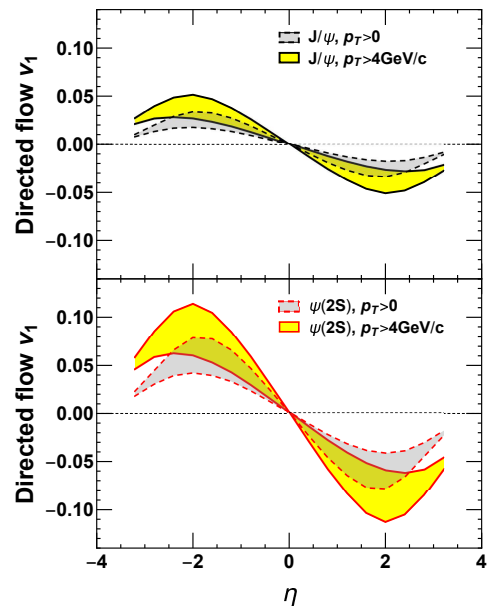}
\caption{(Color online)  
$J/\psi$ (upper panel) and $\psi(2S)$ (lower panel) directed 
flows as a function of rapidity 
$\eta$ with the $p_T$ range $p_T>0$ and $p_T>4$ GeV/c respectively. 
Upper and lower limits of the bands 
corresponds to $\eta_T=2$ and $\eta_T=4$ respectively, 
which is the parameter for the tilt of the initial QGP. 
Other parameters are the same with 
Fig.\ref{labfig-v1}. 
}
\hspace{-0.1mm}
\label{labfig-rangev1}
\end{figure}

In order to check the charmonium directed flows with different tilted QGP, the parameter for 
the tilt in initial longitudinal distribution of energy density 
is shifted to be $\eta_T=2$ 
(for larger tilted medium) and $\eta_T=4$ (for smaller tilted medium). Both $J/\psi$ and 
$\psi(2S)$ directed flows are presented in Fig.\ref{labfig-rangev1}. Their directed flows 
in high $p_T$ bin is more sensitive to the tilted effect. 
In the forward and backward rapidity $|\eta|\sim 2$, prompt $J/\psi$ and prompt $\psi(2S)$ 
directed flows can reach 0.05 and 0.11 respectively at $p_T>4$ GeV/c, 
which are much larger than the 
situations of light charged hadrons $v_1^{\rm charged}\sim 0.005$. 
In the medium with symmetric 
longitudinal distribution, charmonium directed flows drops to zero. 
With the QGP initial conditions given in Fig.\ref{lab-fig1}, charmonium directed flows from 
Fig.\ref{labfig-v1} are close to the lower limits of 
the bands in Fig.\ref{labfig-rangev1}. 
These clear signals 
of charmonium directed flows are the promising probes for the QGP initial tilted distribution 
in forward-backward rapidity in AA collisions. Confirming and measuring the charmonium 
directed flows can help to determine the magnitude of the tilt in initial QGP. 
In Fig.\ref{lab-fig1}-\ref{labfig-rangev1}, 
Charmonium theoretical results 
are in the centrality 0-80\% with 
impact parameter $b=8.3$ fm. 
In other centralities, the tilt of QGP initial energy density 
and the value of $\eta_T$ changes with the colliding centrality. For example, in the 
most central collisions with b=0, the tilt of QGP will disappear.  
In higher colliding energies such as LHC, most charmonia are from 
the regeneration instead of the initial production. The uncertainty of charm quark 
kinetic thermalization in the expanding QGP 
makes directed flows of regenerated charmonia 
ambiguous. The tilt of QGP initial energy density 
in LHC Pb-Pb collisions is also smaller than 
the situation in RHIC Au-Au collisions. Therefore, charmonium directed flows have a 
better chance to be measured at RHIC energies than the LHC energies.  
However, with the 
limited data about charmonium currently available at RHIC energies, 
the $p_T$-integrated $J/\psi$ observables in Fig.\ref{labfig-v1} seems to be 
more prospective.

In this work, we have applied the transport model to study charmonium 
directed flows in the tilted QGP with the symmetry-breaking longitudinal distribution 
produced in semi-central Au+Au collisions at $\sqrt{s_{NN}}=200$ GeV. 
In this centrality, 
charmonium production are dominated by the process of 
parton hard scatterings before the formation 
of QGP. 
In the early stage of QGP evolutions where its tilted shape is more clear, 
charmonium suffers stronger dissociations in the bulk medium with 
high temperatures. They experience biased dissociations due to the 
path-length-difference inside QGP in the 
transverse plane to obtain directed flows, and move quickly out of the bulk medium to keep 
the early information of the initial tilted source.
This makes $J/\psi$ and $\psi(2S)$ directed flows 
several times larger 
than the charged hadrons. The median values of $J/\psi$ and $\psi(2S)$    
directed flows in $p_T>4$ GeV/c can reach $0.03$ and $0.07$ 
in the backward rapidity, which are around 50\% larger than the situation in $p_T>0$. 
The development of charmonium $v_1$ from the path-length-difference 
is different from the situations of light 
hadrons or D mesons which come from QGP anisotropic expansions and usually satisfy the 
mass ordering. 
The signal of charmonium $v_1$ can help to 
extract the early asymmetric deposition of QGP energy density in the forward-backward 
rapidity in symmetric AA collisions.

\vspace{0.5cm}
{\bf Acknowledgement: } This work is supported by NSFC Grant No. 11705125 
and Sino-Germany (CSC-DAAD) Postdoc Scholarship.


\begin{thebibliography}{20}


\bibitem{Liang:2004ph} 
  Z.~T.~Liang and X.~N.~Wang,
  Phys.\ Rev.\ Lett.\  {\bf 94}, 102301 (2005)
  Erratum: [Phys.\ Rev.\ Lett.\  {\bf 96}, 039901 (2006)]

\bibitem{Pang:2016igs} 
  L.~G.~Pang, H.~Petersen, Q.~Wang and X.~N.~Wang,
  Phys.\ Rev.\ Lett.\  {\bf 117}, no. 19, 192301 (2016)
 
\bibitem{Becattini:2015ska} 
  F.~Becattini {\it et al.},
  Eur.\ Phys.\ J.\ C {\bf 75}, no. 9, 406 (2015)
  Erratum: [Eur.\ Phys.\ J.\ C {\bf 78}, no. 5, 354 (2018)].


\bibitem{Adamczyk:2014ipa} 
  L.~Adamczyk {\it et al.} [STAR Collaboration],
  Phys.\ Rev.\ Lett.\  {\bf 112}, no. 16, 162301 (2014)
\bibitem{Bozek:2010bi} 
  P.~Bozek and I.~Wyskiel,
  Phys.\ Rev.\ C {\bf 81}, 054902 (2010).
 
\bibitem{Nara:2016phs} 
  Y.~Nara, H.~Niemi, A.~Ohnishi and H.~Stöcker,
  Phys.\ Rev.\ C {\bf 94}, no. 3, 034906 (2016)


\bibitem{Adam:2019wnk} 
  J.~Adam {\it et al.} [STAR Collaboration],
  arXiv:1905.02052 [nucl-ex].

 \bibitem{Chatterjee:2017ahy} 
  S.~Chatterjee and P.~Bożek,
  Phys.\ Rev.\ Lett.\  {\bf 120}, no. 19, 192301 (2018)

 \bibitem{Dong:2019unq} 
  X.~Dong and V.~Greco,
  Prog.\ Part.\ Nucl.\ Phys.\  {\bf 104}, 97 (2019)


\bibitem{Matsui:1986dk} 
  T.~Matsui and H.~Satz,
  Phys.\ Lett.\ B {\bf 178}, 416 (1986)

\bibitem{Grandchamp:2003uw} 
  L.~Grandchamp, R.~Rapp and G.~E.~Brown,
  Phys.\ Rev.\ Lett.\  {\bf 92}, 212301 (2004)

\bibitem{Zhao:2007hh} 
  X.~Zhao and R.~Rapp,
  Phys.\ Lett.\ B {\bf 664}, 253 (2008)

\bibitem{Zhu:2004nw} 
  X.~l.~Zhu, P.~f.~Zhuang and N.~Xu,
  Phys.\ Lett.\ B {\bf 607}, 107 (2005)

\bibitem{Yan:2006ve} 
  L.~Yan, P.~Zhuang and N.~Xu,
  Phys.\ Rev.\ Lett.\  {\bf 97}, 232301 (2006)

\bibitem{Yao:2018nmy} 
  X.~Yao and T.~Mehen,
  Phys.\ Rev.\ D {\bf 99}, no. 9, 096028 (2019)

\bibitem{Bhanot:1979vb} 
  G.~Bhanot and M.~E.~Peskin,
  Nucl.\ Phys.\ B {\bf 156}, 391 (1979).

\bibitem{Satz:2005hx} 
  H.~Satz,
  J.\ Phys.\ G {\bf 32}, R25 (2006)


\bibitem{Chen:2018kfo} 
  B.~Chen,
  Chinese Physics C  Vol.43, No.12 (2019) 124101.


\bibitem{Andronic:2006ky} 
  A.~Andronic, P.~Braun-Munzinger, K.~Redlich and J.~Stachel,
  Nucl.\ Phys.\ A {\bf 789}, 334 (2007)

\bibitem{Yao:2018sgn} 
  X.~Yao and B.~Müller,
  Phys.\ Rev.\ D {\bf 100}, no. 1, 014008 (2019)

\bibitem{Shi:2017qep} 
  W.~Shi, W.~Zha and B.~Chen,
  Phys.\ Lett.\ B {\bf 777}, 399 (2018)

\bibitem{Liu:2009nb} 
  Y.~p.~Liu, Z.~Qu, N.~Xu and P.~f.~Zhuang,
  Phys.\ Lett.\ B {\bf 678}, 72 (2009)


\bibitem{Du:2015wha} 
  X.~Du and R.~Rapp,
  Nucl.\ Phys.\ A {\bf 943}, 147 (2015)

\bibitem{Lansberg:2016deg} 
  J.~P.~Lansberg and H.~S.~Shao,
  Eur.\ Phys.\ J.\ C {\bf 77}, no. 1, 1 (2017)

\bibitem{Kusina:2017gkz} 
  A.~Kusina, J.~P.~Lansberg, I.~Schienbein and H.~S.~Shao,
  Phys.\ Rev.\ Lett.\  {\bf 121}, no. 5, 052004 (2018)

\bibitem{Emelyanov:1999pkc} 
  V.~Emel'yanov, A.~Khodinov, S.~R.~Klein and R.~Vogt,
  Phys.\ Rev.\ C {\bf 61}, 044904 (2000)
\bibitem{Emelyanov:1998phs} 
  V.~Emel'yanov, A.~Khodinov, S.~R.~Klein and R.~Vogt,
  Phys.\ Rev.\ C {\bf 59}, 1860 (1999)
\bibitem{McGlinchey:2012bp} 
  D.~C.~McGlinchey, A.~D.~Frawley and R.~Vogt,
  Phys.\ Rev.\ C {\bf 87}, no. 5, 054910 (2013)

\bibitem{Chen:2016dke} 
  B.~Chen, T.~Guo, Y.~Liu and P.~Zhuang,
  Phys.\ Lett.\ B {\bf 765}, 323 (2017)


\bibitem{Eskola:2009uj} 
  K.~J.~Eskola, H.~Paukkunen and C.~A.~Salgado,
  JHEP {\bf 0904}, 065 (2009)

\bibitem{Chen:2015iga} 
  B.~Chen,
  Phys.\ Rev.\ C {\bf 93}, no. 5, 054905 (2016)
  
 


\bibitem{Adare:2006kf} 
  A.~Adare {\it et al.} [PHENIX Collaboration],
  Phys.\ Rev.\ Lett.\  {\bf 98}, 232002 (2007)


\bibitem{Schenke:2010nt} 
  B.~Schenke, S.~Jeon and C.~Gale,
  Phys.\ Rev.\ C {\bf 82}, 014903 (2010)
\bibitem{Schenke:2010rr} 
  B.~Schenke, S.~Jeon and C.~Gale,
  Phys.\ Rev.\ Lett.\  {\bf 106}, 042301 (2011)

\bibitem{Huovinen:2009yb} 
  P.~Huovinen and P.~Petreczky,
  Nucl.\ Phys.\ A {\bf 837}, 26 (2010)

\bibitem{Zhou:2014kka} 
  K.~Zhou, N.~Xu, Z.~Xu and P.~Zhuang,
  Phys.\ Rev.\ C {\bf 89}, no. 5, 054911 (2014)
\bibitem{Abelev:2008jga} 
  B.~I.~Abelev {\it et al.} [STAR Collaboration],
  Phys.\ Rev.\ Lett.\  {\bf 101}, 252301 (2008)




\end{thebibliography}
\end{document}